\documentclass[aip,jcp,amsmath,amssymb,reprint,floatfix]{revtex4-1}

\usepackage{graphicx}
\usepackage{dcolumn}
\usepackage{bm}

\usepackage{hyphenat}

\usepackage{multirow}
\usepackage{booktabs}

\usepackage{amsmath}
\usepackage{amsfonts}
\usepackage{amssymb}
\usepackage{amsbsy}
\usepackage{mathrsfs}
\usepackage{upgreek}
\usepackage[cbgreek]{textgreek}

\newcommand{\Bra}[1]{\ensuremath{\bigl\langle {#1} \bigl\lvert}}
\newcommand{\Ket}[1]{\ensuremath{\bigr\rvert {#1} \bigr\rangle}}
\newcommand{\BraKet}[2]{\ensuremath{\bigl\langle {#1} \bigl\lvert {#2} \bigr\rangle}}
\newcommand{\tBraKet}[3]{\ensuremath{\bigl\langle {#1} \bigl\lvert {#2} \bigl\lvert {#3} \bigr\rangle}}
\graphicspath{{./figures/}}

\begin{document}

\title{Ab Initio Effective One-Electron Potential Operators. II.
Applications for Exchange-Repulsion Energy in Effective Fragment Potentials}

\author{Bartosz B{\l}asiak}
\email[]{blasiak.bartosz@gmail.com}
\homepage[]{https://www.polonez.pwr.edu.pl}

\author{Robert W. G{\'o}ra}
\author{Wojciech Bartkowiak}

\affiliation{Department of Physical and Quantum Chemistry, Faculty of Chemistry, 
Wroc{\l}aw University of Science and Technology, 
Wybrze{\.z}e Wyspia{\'n}skiego 27, Wroc{\l}aw 50-370, Poland}

\date{\today}

\begin{abstract}
In Paper~I, the effective one\hyp{}electron
potentials (OEP) method was introduced and demonstrated as an
efficient approach
to reduce the computational cost
of evaluation of the charge-transfer interaction energy
within the effective fragment potential method (EFP2)
by an average factor of 20, making it no longer a bottleneck
in EFP2\hyp{}based simulations of complex systems.
Here, the OEP technique 
is used to enhance computational efficiency in evaluating the
exchange\hyp{}repulsion EFP2 interaction energy
by redefining the first\hyp{}order repulsive term of 
Murrell~et al. [Murrell et al., Proc. R. Soc. Lond. A 284, 566 (1965)] through
the extended density fitting in incomplete auxiliary basis.
In the proposed approach the evaluation of the kinetic energy integrals is no
longer required and the computational cost can be reduced roughly by a factor
of 1.5 as compared to the original EFP2 formulation.
\end{abstract}

\pacs{}

\maketitle


\section{\label{s:1.introduction}Introduction}

The exchange\hyp{}repulsion (EXR) interaction energy 
usually describes the non\hyp{}electrostatic repulsion
between two isolated and unperturbed 
wavefunctions in an interacting
molecular 
complex.\cite{Murrell.Randic.Williams.Longuet-Higgins.ProcRSocLondA.1965,Otto.Ladik.ChemPhys.1975,Hayes.Stone.MolPhys.1984,Jeziorski.Moszynski.Szalewicz.ChemRev.1994,Jensen.Gordon.MolPhys.1996,Hesselmann.Jansen.Schutz.JCP.2005,Mandado.Hermida-Ramon.JCTC.2011}
In the
second generation of the
effective fragment potentials (EFP2) method\cite{Gordon.Smith.Xu.Slipchenko.AnnuRevPhysChem.2013,
   Nguyen.Pachter.Day.JCP.2014,
   Day.Jensen.Gordon.Webb.Stevens.Krauss.Garmer.Basch.Cohen.JCP.1996}
the EXR energy is an important component of the total interaction energy.
It is
derived from the intermolecular perturbation theory
of Murrell et al.\cite{Murrell.Randic.Williams.Longuet-Higgins.ProcRSocLondA.1965,Otto.Ladik.ChemPhys.1975} 
up to the second order with respect to the wavefunction overlap
assuming Hartree\hyp{}Fock\cite{Roothaan.RevModPhys.1951} (HF) wavefunctions,
and is given by~\cite{Jensen.JCP.1996,Jensen.Gordon.MolPhys.1996,Jensen.Gordon.JCP.1998}
\begin{equation} \label{e:exr-efp2}
 E^{\rm Ex-Rep} \approx
 E^{\rm Ex} +
 E^{\rm Rep}(S^{-1}) + 
 E^{\rm Rep}(S^{-2}) \;.
\end{equation}
In the above equation,
$E^{\rm Ex}$ is the exchange energy,\cite{Jensen.JCP.1996}
\begin{equation} \label{e:exc-efp2}
 E^{\rm Ex} \approx 2 \sum_{i\in A}^{\rm Locc} \sum_{j\in B}^{\rm Locc} 
 \sqrt{\frac{-2\ln{\vert S_{ij}\vert}}{\pi}} \frac{S^2_{ij}}{r_{ij}} \;,
\end{equation}
whereas the first\hyp{} and second\hyp{}order repulsion terms
are accordingly~\cite{Jensen.Gordon.MolPhys.1996,Jensen.Gordon.JCP.1998}
\begin{multline} \label{e:rep-efp2.s1}
 E^{\rm Rep}(S^{-1}) \approx - 2 \sum_{i\in A}^{\rm Locc} \sum_{j\in B}^{\rm Locc} S_{ij} 
 \\\times 
 \Bigg\{
 \sum_{k\in A}^{\rm Locc} F_{ik}^A S_{kj} 
+\sum_{l\in B}^{\rm Locc} F_{jl}^A S_{li} 
 -2T_{ij}
 \Bigg\}
\end{multline}
and 
\begin{multline} \label{e:rep-efp2.s2}
 E^{\rm Rep}(S^{-2}) \approx 2 \sum_{i\in A}^{\rm Locc} \sum_{j\in B}^{\rm Locc} S_{ij}^2 
 \\ \times 
 \Bigg\{
 \sum_{x\in A}^{\rm At}  \frac{-Z_x}{r_{xb}}
+\sum_{y\in B}^{\rm At}  \frac{-Z_y}{r_{ya}} 
+\sum_{k\in A}^{\rm Locc} \frac{   2}{r_{jk}}
+\sum_{l\in B}^{\rm Locc} \frac{   2}{r_{il}}
-r_{ij}^{-1}
 \Bigg\} \;.
\end{multline}
The indices $i$, $j$, $k$ and $l$ label the localized occupied orbitals (LMO's, denoted by `Locc')
located at their charge centroids ${\bf r}_{i(j)}$,
whereas $x$ and $y$ label atomic nuclei (denoted by `At') with atomic numbers 
$Z_{x(y)}$ located at ${\bf r}_{x(y)}$. The relative distances are
defined by $r_{uw}=\vert {\bf r}_u-{\bf r}_w\vert$.
$S_{ij}$ and $T_{ij}$ are the overlap and kinetic\hyp{}energy
integrals, respectively, whereas $F_{ik}^X$ are the Fock matrix elements
of unperturbed monomer $X$, all defined in LMO basis.
It is known that evaluating Eq.~\eqref{e:exr-efp2}
is remarkably efficient and gives usually very accurate
estimates of the reference EXR energies according to the symmetry 
adapted perturbation theory\cite{Jeziorski.Moszynski.Szalewicz.ChemRev.1994} (SAPT).
The computational cost to evaluate
the exchange and second\hyp{}order terms
is much smaller than that of the first\hyp{}order term,
which requires evaluation of not only the one\hyp{}electron overlap,
but also the kinetic
energy integrals in a space of occupied molecular orbital (MO) basis.

In this work, an alternative formulation for the first\hyp{}order
term is proposed based on the effective one\hyp{}electron potential (OEP)
operator technique developed in the preceding contribution.~\cite{Blasiak.Bednarska.Choluj.Bartkowiak.JCP.2019}
It is shown that using this approach calculation of kinetic energy integrals can be avoided
what leads to increased computational efficiency.

\section{Results}

\begin{figure}[t]
\includegraphics[width=0.44\textwidth]{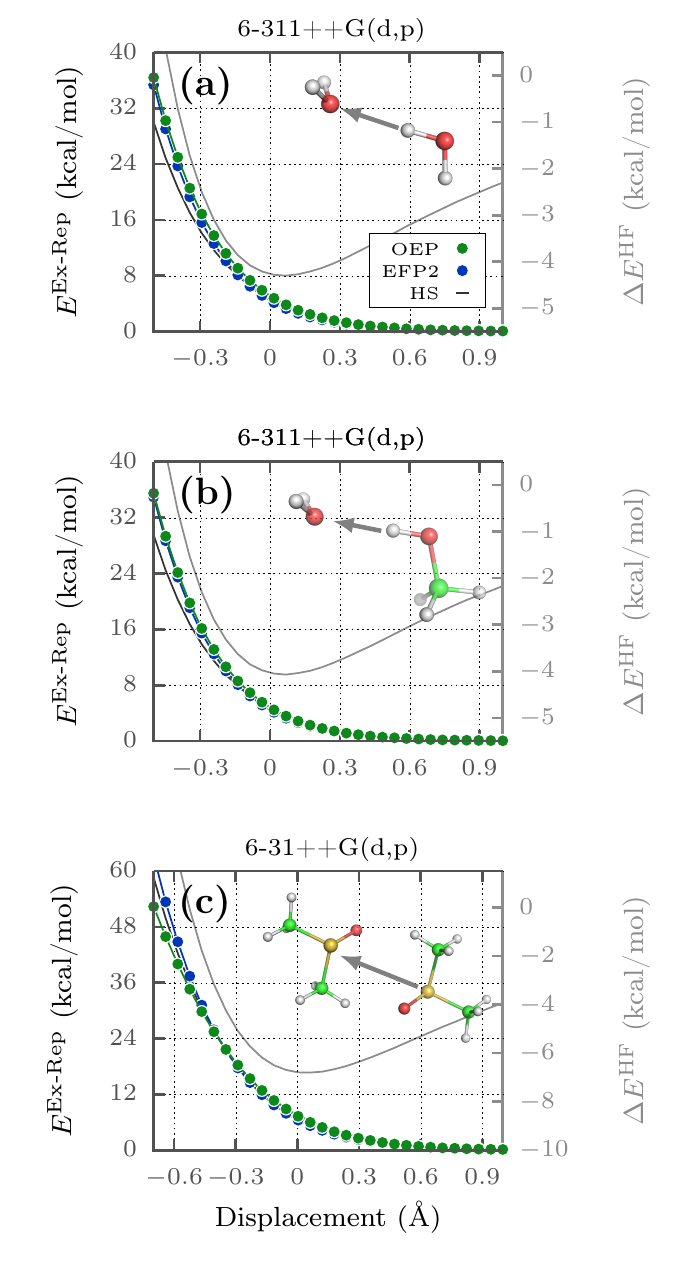}
\caption{\label{f:fig-1} {\bf Asymptotic dependence of the exchange\hyp{}repulsion interaction energy
for selected bi\hyp{}molecular complexes.} 
(a) water dimer, 
(b) water\hyp{}methanol complex, and 
(c) DMSO dimer in $S_i$ symmetry,
were one molecule has been translated
along the vector specified in the insets relative to initial geometry,
optimized at HF/6-31+G(d,p) level of theory.
The 
total interaction energy
is also shown for comparison in light grey color in this figure.
Interaction energies were obtained by using the 6-311++G(d,p)
primary basis set for (a) and (b), and 6-31++G(d,p) for (c).
In OEP calculations, EDF-1 scheme combined with small optimized auxiliary basis sets
(cf. Supplementary Information) were used.
} 
\end{figure}

A theoretical foundation of the
first\hyp{}order repulsive term in the EFP2 model
is the perturbation theory of
Murrell et~al.~\cite{Murrell.Randic.Williams.Longuet-Higgins.ProcRSocLondA.1965} 
in which
\begin{multline} \label{e:rep.murrell-etal.S1}
    E^{\rm Rep}(S^{-1}) = -2\sum_{i\in A}^{\rm Locc} \sum_{j\in B}^{\rm Locc}
               S_{ij} \Big\{
         - W^A_{ij} 
         - W^B_{ij} \\
 + \sum_{k\in A}^{\rm Occ} \left[ 2\BraKet{ij}{kk} - \BraKet{ik}{jk} \right] 
 + \sum_{l\in B}^{\rm Occ} \left[ 2\BraKet{ij}{ll} - \BraKet{il}{jl} \right]
                \Big\}
\end{multline}
where $W^A_{ij} \equiv \tBraKet{i}{\hat{v}^A_{\rm nuc}}{j}$,
$\hat{v}^A_{\rm nuc}$
is the electrostatic potential operator due to nuclei
and
ERIs are defined by
\begin{equation} \label{e:eri}
	\BraKet{\alpha\beta}{\gamma\delta} \equiv
	\iint 
	\frac{ \phi_\alpha^{*}({\bf r}_1) \phi_\beta({\bf r}_1) 
	       \phi_\gamma^{*}({\bf r}_2) \phi_\delta({\bf r}_2) }{ \vert {\bf r}_1 - {\bf r}_2 \vert}
	d{\bf r}_1 d{\bf r}_2  \;.
\end{equation}
Now, note that the Coulomb and exchange integrals can be re\hyp{}cast as follows:
\begin{subequations}
 \begin{align}
 \BraKet{ij}{kk} &\equiv - \sum_{\mu\in A} 
     C_{\mu i}^A \tBraKet{\mu}{\hat{v}_{kk}}{j} \;, \\
 \BraKet{ik}{jk} &\equiv - \sum_{\mu\in A} 
     C_{\mu k}^A \tBraKet{\mu}{\hat{v}_{ik}}{j} \;.
 \end{align}
\end{subequations}
%
%
%
In the above equations, the auxiliary potential operators are given by
\begin{equation}
  \hat{v}_{ik} \equiv - \int d{\bf r} \Ket{{\bf r}} 
        \left[
        \int d{\bf r}' \frac{\phi_i^{*}({\bf r}') \phi_k({\bf r}')}{\vert {\bf r}' - {\bf r}\vert}
        \right] \Bra{{\bf r}} \;.
\end{equation}
ERIs can be effectively eliminated from Eq.~\eqref{e:rep.murrell-etal.S1} 
by using the following prescription~\cite{Blasiak.Bednarska.Choluj.Bartkowiak.JCP.2019}
\begin{equation} \label{e:ft-reduction}
	\sum_t \sum_{ij}\sum_{kl\in A} {\mathcal{F}}_t\left[ 
   \BraKet{ij}{k^Al^A}
 \right] = \sum_{ij} \tBraKet{i}{\hat{v}_{\text{eff}}^A}{j} \;.
\end{equation}
where ${\mathcal{F}}_t$ is a certain and well defined functional of ERIs.
Application of Eq.~\eqref{e:ft-reduction} enables to define
a joint OEP
operator constructed from nuclear, Coulomb and exchange
contributions as
\begin{equation}
 -W^A_{ij} + 
 \sum_{k\in A}^{\rm Occ} 
  \left\{ 2\BraKet{ij}{kk} - \BraKet{ik}{jk} \right\}
\equiv \sum_{\mu\in A} \tBraKet{\mu}{ 
\hat{v}^{A[\mu i]}_{\rm eff}
 }{j}
\end{equation}
with
\begin{equation}
 \hat{v}^{A[\mu i]}_{\rm eff} \equiv C_{\mu i} \hat{v}^A_{\rm nuc} + 
 \sum_{k\in A}^{\rm Occ} \left[
 2C_{\mu i}^A \hat{v}^A_{kk} - C_{\mu k}^A \hat{v}^A_{ik}
 \right] \;.
\end{equation}
%
%
%
{
\renewcommand{\arraystretch}{1.4}
\begin{table*}[t]
\caption[]
{{\bf CPU timings in miliseconds of exchange\hyp{}repulsion single point 
energy calculations via various methods\footnotemark[1]}
}
\label{t:tab-1}
\begin{ruledtabular}
\begin{tabular}{llccccccccc}
                      && \multicolumn{2}{c}{(H$_2$O)$_2$\footnotemark[2]} 
                      && \multicolumn{2}{c}{H$_2$O--HOCH$_3$\footnotemark[2]} 
                      && \multicolumn{2}{c}{(DMSO)$_2$\footnotemark[3]} \\
\cline{3-4}
\cline{6-7}
\cline{9-10}
Hayes \& Stone\footnotemark[4] && 1.16$\times 10^3$  & (4.81)  
                               && 1.12$\times 10^4$  & (4.72)
                               && 1.31$\times 10^5$  & (7.34) \\
Murrell et al.\footnotemark[5] && 2.33$\times 10^2$  & (4.53)  
                               && 3.87$\times 10^3$  & (4.44)
                               && 2.79$\times 10^4$  & (6.84) \\
EFP2                           && 0.496              & (4.45)  
                               && 0.84               & (4.44)
                               && 4.65   & (6.56) \\
OEP/large\footnotemark[6]      && 0.884  & (4.83)  
                               && 1.836  & (4.83)
                               && 8.95   & (6.77) \\
OEP/mini\footnotemark[6]       && 0.324  & (5.13)  
                               && 0.566  & (4.81)
                               && 3.00   & (7.41) \\
\end{tabular}
\end{ruledtabular}
\footnotetext[1]{1.2 GHz AMD EPYC\texttrademark{} 7301 16-Core Processor, calculations performed on 1 core. 
Exchange\hyp{}repulsion energies are given in parentheses for reference (kcal/mol).
Time profiling of the code performance was carried out as in Ref.~\cite{Blasiak.Bednarska.Choluj.Bartkowiak.JCP.2019}.
See also the implementation details in the main text.}
\footnotetext[2]{Primary basis set: 6-311++G(d,p).}
\footnotetext[3]{Primary basis set: 6-31++G(d,p).}
\footnotetext[4]{Reference~\cite{Hayes.Stone.MolPhys.1984}.}
\footnotetext[5]{Reference~\cite{Murrell.Randic.Williams.Longuet-Higgins.ProcRSocLondA.1965}.}
\footnotetext[6]{Same as EFP2 but with $E^{\rm Rep}(S^{-1})$ replaced by formula in Eq.~\eqref{e:rep.murrell-etal.S1.oep}. 
Auxiliary basis sets for the EDF in $E^{\rm Rep}(S^{-1})$ 
from Eq.~\eqref{e:rep.murrell-etal.S1.oep}
are: `large' - aug-cc-pVDZ-jkfit for (H$_2$O)$_2$ and H$_2$O--HOCH$_3$ systems as well as
aug-cc-pVDZ-ri for (DMSO)$_2$ system;
`mini' - optimized small basis sets (see Supporting Information).}
\end{table*}
}
\\ On the other hand, it immediately follows that
\begin{equation}
 \hat{v}^{A[\mu i]}_{\rm eff} \Ket{\mu} = 
  \sum_{k\in A}^{\rm Occ} \left\{
     2\hat{v}^A_{kk} \Ket{i} - \hat{v}^A_{ik} \Ket{k}
  \right\} \;.
\end{equation}
For practical calculations, the right hand side of the above equation can be expanded 
in the auxiliary basis,
\begin{equation} \label{e:v-oep.rep}
  \sum_{k\in A}^{\rm Occ} \left\{
     2\hat{v}^A_{kk} \Ket{i} - \hat{v}^A_{ik} \Ket{k}
  \right\} \cong
  \sum_{\xi\in A}^{\rm DF} 
  V_{\xi i}^A \Ket{\xi} \;,
\end{equation}
where the matrix ${\bf V}^A$ can be considered as a set of effective fragment parameters.
Doing the same operations on the twin operators associated with the molecule $B$
original theory of Murrell et. al from Eq.~\eqref{e:rep.murrell-etal.S1}
reduces to
\begin{equation} \label{e:rep.murrell-etal.S1.oep}
    E^{\rm Rep}(S^{-1}) \cong 
 -2\sum_{i\in A}^{\rm Occ} \sum_{j\in B}^{\rm Occ}
               S_{ij} \Big\{
           \sum_{\xi \in A}^{\rm DF} V_{\xi i}^A S_{\xi j}
         + \sum_{\eta\in B}^{\rm DF} V_{\eta j}^B S_{\eta i}
                \Big\} \;,
\end{equation}
where the
effective fragment parameters
can be obtained 
from the extended density fitting\cite{Blasiak.Bednarska.Choluj.Bartkowiak.JCP.2019} (EDF)
by
\begin{equation} \label{e:rep.murrell-etal.S1.oep-edf}
            V_{\xi i}^X = \sum_{\eta\in X}^{\rm DF} 
                          \sum_{\varepsilon,\zeta\in X}^{\rm RI}
                          \left[ {\bf R}^{-1} \right]_{\xi\eta} R_{\eta\varepsilon} 
                          \left[ {\bf S}^{-1} \right]_{\varepsilon\zeta} 
                          \tBraKet{\zeta}{\hat{v}_{\rm eff}^{X[\zeta i]}}{i}
                \;.
\end{equation}
In this procedure, $R_{\alpha\gamma}= \iint d{\bf r}_1 d{\bf r}_2 \frac{\phi_{\alpha}^*({\bf r}_1) \phi_{\beta}({\bf r}_2) }{\lvert {\bf r}_1 - {\bf r}_2 \rvert}$, ${\bf S}$ is the matrix of overlap integrals in AO basis
and the required OEP matrix elements can be calculated from
\begin{multline} \label{e:rep.murrell-etal.S1.oep-v}
   \tBraKet{\zeta}{\hat{v}_{\rm eff}^{X[\zeta i]}}{i}
     = -\sum_{x\in X}^{\rm At} W_{\zeta i}^{(x)} \\
        + \sum_{\beta\gamma\delta\in X}^{\rm AO}
           \left\{ 
             2 C_{\beta i}^X D_{\gamma\delta}^X - C_{\gamma i}^X D_{\beta \delta}^X
           \right\}
           \BraKet{\zeta\beta}{\gamma\delta} \;,
\end{multline}
where ${\bf D}^X$ and ${\bf C}^X$ are the one\hyp{}particle AO density and
the LCAO\hyp{}MO matrices of isolated molecule $X$, respectively.
%
Eq.~\eqref{e:rep.murrell-etal.S1.oep} and Eq.~\eqref{e:rep-efp2.s1} have almost
the same form, with two exceptions in the new OEP formulation: 
(i) only overlap integrals are needed, which are relatively easy to
compute. Furthermore, there is no need to evaluate the kinetic
energy integrals, as in the original EFP2
formulation, which are computationally slightly more demanding;
(ii) overlap integrals need to be evaluated also between
auxiliary basis set. The smaller the size of the auxiliary basis, the less expensive
evaluation of the $E^{\rm Rep}(S^{-1})$ becomes. 
The alternative EFP2 formulation 
is therefore still given by Eq.~\eqref{e:exr-efp2}
%
%
but with $E^{\rm Rep}(S^{-1})$ replaced by OEP\hyp{}based
first\hyp{}order term from Eq.~\eqref{e:rep.murrell-etal.S1.oep}.

In our in\hyp{}house plug-in to the {\sc Psi4}
quantum chemistry
program,\cite{Psi4.JCTC.2017}
we implemented EXR energy models from Eq.~\eqref{e:exr-efp2}
(EFP2), Eq.~\eqref{e:rep.murrell-etal.S1.oep} (OEP), as well as 
the approximate intermolecular perturbation theory with exchange of 
Murrell et~al.\cite{Murrell.Randic.Williams.Longuet-Higgins.ProcRSocLondA.1965},
and the exact EXR energy of Hayes \& Stone\cite{Hayes.Stone.MolPhys.1984},
here referred to as the HS model.
The latter was
implemented in the dimer\hyp{}centered basis set\cite{Chalasinski.Gutowski.MolPhys.1985} (DCBS),
to eliminate the basis set
superposition error (BSSE) in the benchmark calculations.
The Boys method\cite{Boys.RevModPhys.1960} was used to localize
molecular orbitals---the same as in the original EFP2 formulation.
Note here that, since
Murrell et~al. theory is invariant with respect to unitary transformation
of molecular orbitals, the OEP\hyp{}based expression 
in Eq.~\eqref{e:rep.murrell-etal.S1.oep} is also invariant and does not require
orbital localization. 
%
%
%
\begin{figure*}[t]
\includegraphics[width=0.8\textwidth]{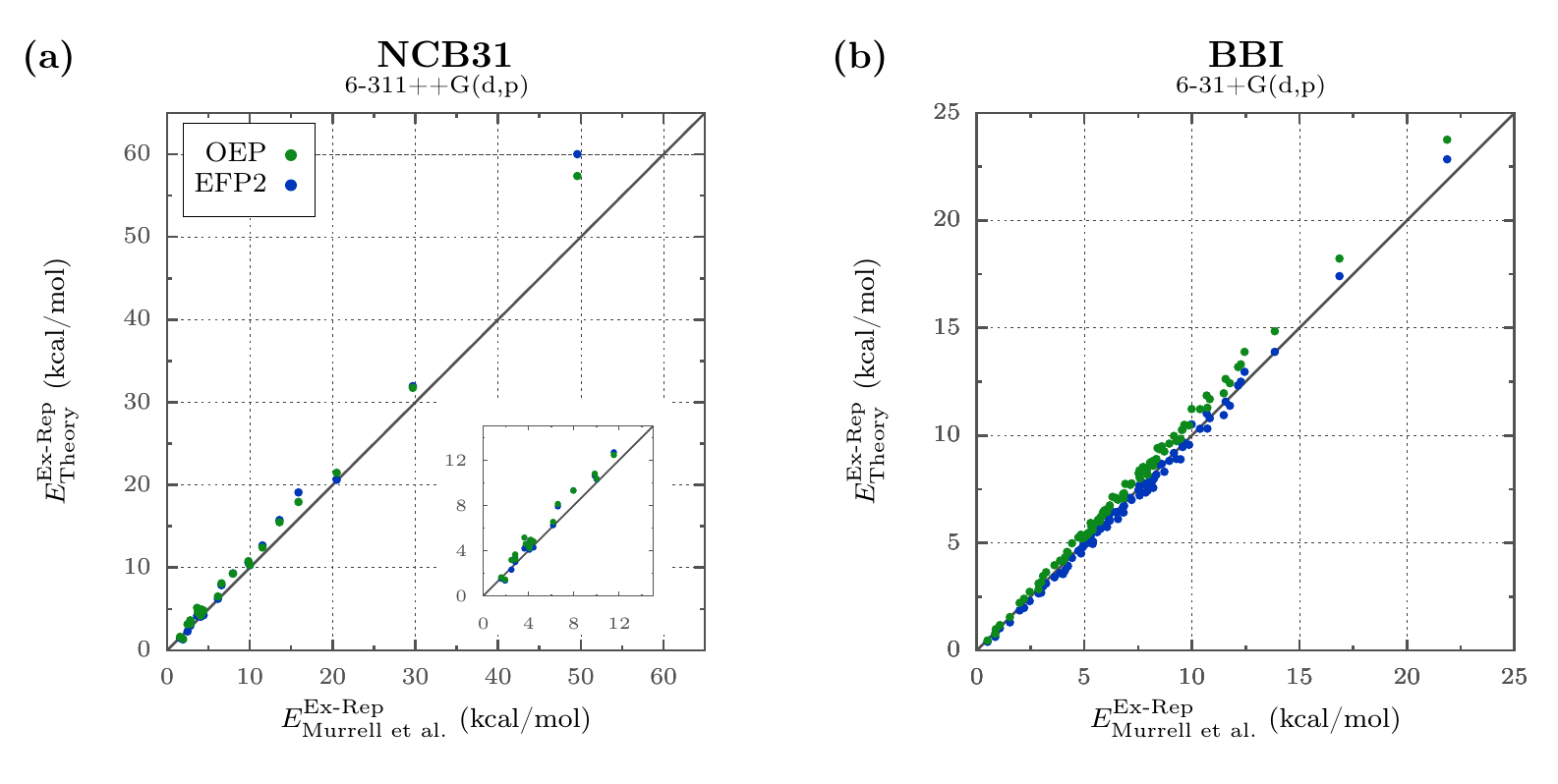}
\caption{\label{f:fig-2} {\bf Accuracy of the OEP and EFP2 models of exchange\hyp{}repulsion energy
across various bi\hyp{}molecular systems.} 
(a) NCB31 
database\cite{Zhao.Schultz.Truhlar.JCTC.2006,
Zhao.Truhlar.JCTC.2005,Zhao.Schultz.Truhlar.JCTC.2006,Zhao.Schultz.Truhlar.JCP.2005} 
of non\hyp{}covalent interactions
and
(b) BBI subset\cite{Burns.Faver.Zheng.Marshall.Smith.Vanommeslaeghe.MacKerell.Merz.Sherrill.JCP.2017} 
of backbone\hyp{}backbone interactions in proteins from the BioFragment Database.
For the OEP calculations, the EDF-1 scheme with the aug-cc-pVDZ-jkfit auxiliary basis set
was used.
} 
\end{figure*}

Three complexes: 
(H$_2$O)$_2$, 
H$_2$O--HOCH$_3$, and
(DMSO)$_2$,
were chosen as model systems to analyze the asymptotic dependence 
of EXR energy, which is shown in Figure~\ref{f:fig-1}. 
The reference (zero\hyp{}displacement) geometries
were obtained as described in Ref.\cite{Blasiak.Bednarska.Choluj.Bartkowiak.JCP.2019} 
and the structures  
along the translation direction are depicted
in the insets in Figure~\ref{f:fig-1}.
Energy\hyp{}optimizations were performed at the HF/6-31+G(d,p) level,
as implemented in 
the {\sc Gaussian16} quantum chemistry program package.\cite{Gaussian16}
OEP and EFP2 models correctly describe the EXR energy at all separations
for the studied model systems with very similar accuracy. 
However,
contrary to our previous application of the OEP technique\cite{Blasiak.Bednarska.Choluj.Bartkowiak.JCP.2019},
where approximately 20\hyp{}fold speedups as compared to the EFP2 model were achieved
for the evaluation of the charge\hyp{}transfer energy,
the CPU timings of EXR calculations in the OEP method are comparable but roughly 2
times higher than that of the EFP2 model
when the usual density fitting auxiliary basis sets such as aug-cc-pVDZ-jkfit or aug-cc-pVDZ-ri are used
(Table~\ref{t:tab-1}).
Computational cost of $E^{\rm Rep}(S^{-1})$ in EFP2 model
is approximately $s(on^2 + o^2n) + t(on^2 + o^2n) + 2o^3$, whereas in OEP model
is $s(on^2 + o^2n) + 2ao^2$, both of which have comparable magnitudes. 
Here, $s$ and $t$ denote the relative costs of evaluation of the overlap
and kinetic-energy one electron integrals (OEIs), whereas $o$, $n$ and $a$ is
the number of LMOs, primary AOs and auxiliary AOs, respectively.
It is clear that the critical parameter is the number of auxiliary basis functions $a$,
which should be comparable with the number of occupied orbitals $o$ 
as $t$ is usually only 2--3 times larger than $s$.
To investigate this issue more thoroughly we developed small auxiliary basis sets for H$_2$O, CH$_3$OH and DMSO, 
via the basis set optimization method described in Appendix~A to Paper I
in conjunction with the basin hopping global optimization algorithm\cite{Wales.EnergyLandscapes.2003,
Wales.Scheraga.Science.1999,
Wales.Doye.JPCA.1997,Li.Scheraga.PNAS.1987}
as implemented in the SciPy Python library\cite{SciPy.2019}
(see Supplementary Information).
Applying such basis sets
reduces computational effort appreciably and makes the OEP method on average 1.5 times faster than EFP2.
For instance, calculation of the EXR energy in the DMSO dimer by using 6-31++G(d,p) primary
basis set and 
1.2~GHz AMD EPYC\texttrademark{} 7301 16-Core Processor
requires about 4.6~ms and 3.0~ms when EFP2 and OEP models are used, respectively.

To investigate the accuracy of the OEP\hyp{}based repulsion
term across
a variety of
interacting systems, a selection of
bi\hyp{}molecular complexes from the non\hyp{}covalent
interactions database NCB31 developed by the Truhlar's 
group,\cite{Zhao.Schultz.Truhlar.JCTC.2006,
Zhao.Truhlar.JCTC.2005,Zhao.Schultz.Truhlar.JCTC.2006,Zhao.Schultz.Truhlar.JCP.2005}
as well as the BioFragment Database subset BBI for backbone\hyp{}backbone
interactions in proteins of Sherrill group,\cite{Burns.Faver.Zheng.Marshall.Smith.Vanommeslaeghe.MacKerell.Merz.Sherrill.JCP.2017} 
as implemented in the {\sc Psi4}
program,\cite{Psi4.JCTC.2017}
was utilized. The computed
EXR energies were compared with the reference Murrell~et~al. results (Figure~\ref{f:fig-2})
as well as with the reference HS results (Figure~S1).
On average, OEP and EFP2 methods are in good agreement
with the reference models and correlation $R^2$ coefficients
are between 90--99\% in all data sets. 
The mutual differences between EFP2 and OEP estimations of 
$E^{\rm Rep}(S^{-1})$ are on the average 0.7 kcal/mol  (Figure~S2) which shows that the ERI elimination proposed here is accurate.
Root mean square errors (RMSE) of EXR energy estimation
via OEP and EFP2 models
are around 2.0~kcal/mol and 2.6~kcal/mol in the NCB31/6-311++G(d,p) data set, and 0.6~kcal/mol 
and 0.3--1.2~kcal/mol
in the BBI/6-31+G(d,p) data set, respectively. However,
while the OEP and EFP2 models tend to overestimate
the EXR energy as compared to the Murrell~et~al. reference
by 7--12\%
(except for the EFP2 model in the BBI/6-31+G(d,p) data set that shows very good agreement), 
they consistently underestimate the EXR energy
as compared to the HS reference by 0.7--7\% (OEP)
and 8--15\% (EFP2).
This is mostly due to BSSE, that is corrected for
only in the reference calculations with the HS model.
Nevertheless,
OEP model is rather of roughly comparable accuracy as EFP2 model in all the systems studied.
Together with the performance data of both models
from Table~\ref{t:tab-1}, OEP model consistently outperforms the
EFP2 approach provided
that the minimal auxiliary basis set is used.

\section{\label{s:6.conclusions}Summary and a few concluding remarks}

In this work, the effective one-electron potential (OEP) technique,
proposed in the preceding paper for the
effective elimination of electron
repulsion integrals in ab initio calculations,
was utilized to reduce the computational cost
of evaluation of the exchange repulsion (EXR) energy in the EFP2 model.
Starting from the first-order formula for EXR in Murrell et~al. perturbation theory, 
being a foundation of the EFP2 EXR term,
OEP\hyp{}based expression was derived which requires evaluation of
only the overlap one-electron integrals, in contrast to EFP2
approach which requires also the kinetic-energy one-electron
integrals. 
The reported results indicate that a following model should be more
efficient for calculations of the 
total exchange\hyp{}repulsion energy in the EFP2 model: 
(i) $E^{\rm Ex}$ and $E^{\rm Rep}(S^{-2})$ are evaluated
from the original EFP2 formulae of Jensen and Gordon, i.e.,
from Eqs.~\eqref{e:exc-efp2} and \eqref{e:rep-efp2.s2}, respectively;
(ii) $E^{\rm Rep}(S^{-1})$ term is evaluated from Eq.~\eqref{e:rep.murrell-etal.S1.oep}
and assuming a small auxiliary basis sets, optimized for each
EFP2 fragment separately.

\begin{acknowledgments}
This project is carried out under POLONEZ programme which has received funding from the European Union's
Horizon~2020 research and innovation programme under the Marie Sk{\l}odowska-Curie grant agreement 
No.~665778. This project is funded by National Science Centre, Poland 
(grant~no. 2016/23/P/ST4/01720) within the POLONEZ 3 fellowship.
Wroclaw Centre for Networking and Supercomputing (WCSS) is acknowledged for
computational resources.
\end{acknowledgments}


\bibliography{references}

\end{document}


\renewcommand{\thefigure}{S\arabic{figure}}
\setcounter{figure}{0}

\title{Supplementary Information\\Ab Initio Effective One-Electron Potential Operators. II.
Applications for Exchange-Repulsion Energy in Effective Fragment Potentials}

\author{Bartosz B{\l}asiak}
\email[]{blasiak.bartosz@gmail.com}
\homepage[]{https://www.polonez.pwr.edu.pl}

\author{Robert W. G{\'o}ra}
\author{Wojciech Bartkowiak}

\affiliation{Department of Physical and Quantum Chemistry, Faculty of Chemistry, 
Wroc{\l}aw University of Science and Technology, 
Wybrze{\.z}e Wyspia{\'n}skiego 27, Wroc{\l}aw 50-370, Poland}

\date{\today}

\pacs{}

\maketitle

\tableofcontents

\section{Minimalistic auxiliary basis sets for water, methanol and DMSO molecules}

The basis sets are given below. Note that in case of DMSO, the basis set size is smaller than the
minimal basis size (3p orbital on S atoms were not necessary to obtain good fitting in the test basis set).

\begin{verbatim}
[ water ]
cartesian
# primary basis set: 6-311++G** 6D
****
H     0
S   1   1.00
        0.500             1.000000
****
O     0
S   2   1.00
      937.182             0.986162 
        6.892             0.014190 
S   1   1.00
       45.156             1.000000
P   2   1.00
       24.823             0.277579 
        2.783             0.301334 
****

[ methanol ]
cartesian
# primary basis set: 6-311++G** 6D
****
H     0
S   1   1.00
        0.539             1.000000
****
C     0
S   1   1.00
      939.683             1.000000
S   1   1.00
       30.185             1.000000
P   1   1.00
        5.832             1.000000
P   1   1.00
        0.326             1.000000
****
O     0
S   1   1.00
      731.638             1.000000
S   1   1.00
       37.387             1.000000
P   1   1.00
        9.214             1.000000
P   1   1.00
        0.368             1.000000
****

[ dmso ]
cartesian
# primary basis set: 6-31++G** 6D
****
H     0
S   1   1.00
        0.365             1.000000
****
C     0
S   1   1.00
      207.472             1.000000
S   1   1.00
       15.052             1.000000
P   1   1.00
        7.228             1.000000
****
O     0
S   1   1.00
      302.339             1.000000
S   1   1.00
       15.248             1.000000
P   1   1.00
        7.612             1.000000
****
S     0
S   1   1.00
      689.986             1.000000
S   1   1.00
      145.419             1.000000
S   1   1.00
        9.046             1.000000
P   1   1.00
       28.979             1.000000
****
\end{verbatim}

\vfill
\section{Accuracy of OEP and EFP2 models for exchange-repulsion energy estimation}
%
\begin{figure*}[h]
\includegraphics[width=0.8\textwidth]{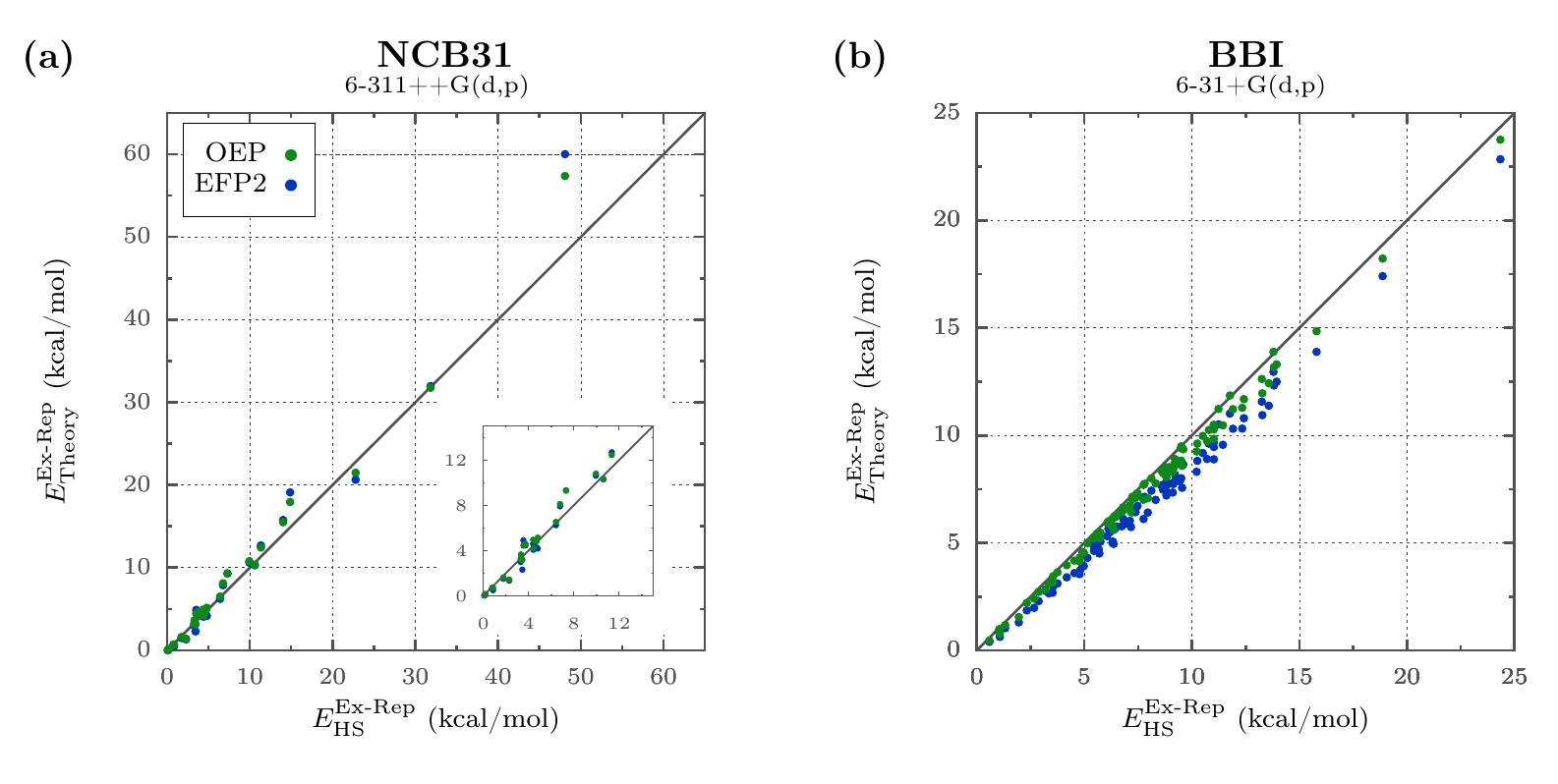}
\caption{\label{f:fig-2} {\bf Accuracy of the OEP and EFP2 models of exchange\hyp{}repulsion energy
across various bi\hyp{}molecular systems.} 
(a) NCB31 
database\cite{Zhao.Schultz.Truhlar.JCTC.2006,
Zhao.Truhlar.JCTC.2005,Zhao.Schultz.Truhlar.JCTC.2006,Zhao.Schultz.Truhlar.JCP.2005} 
of non\hyp{}covalent interactions
and
(b) BBI subset\cite{Burns.Faver.Zheng.Marshall.Smith.Vanommeslaeghe.MacKerell.Merz.Sherrill.JCP.2017} 
of backbone\hyp{}backbone interactions in proteins from the BioFragment Database.
For the OEP calculations, the EDF-1 scheme with the aug-cc-pVDZ-jkfit auxiliary basis set
was used.
} 
\end{figure*}
\begin{figure*}[b]
\includegraphics[width=0.8\textwidth]{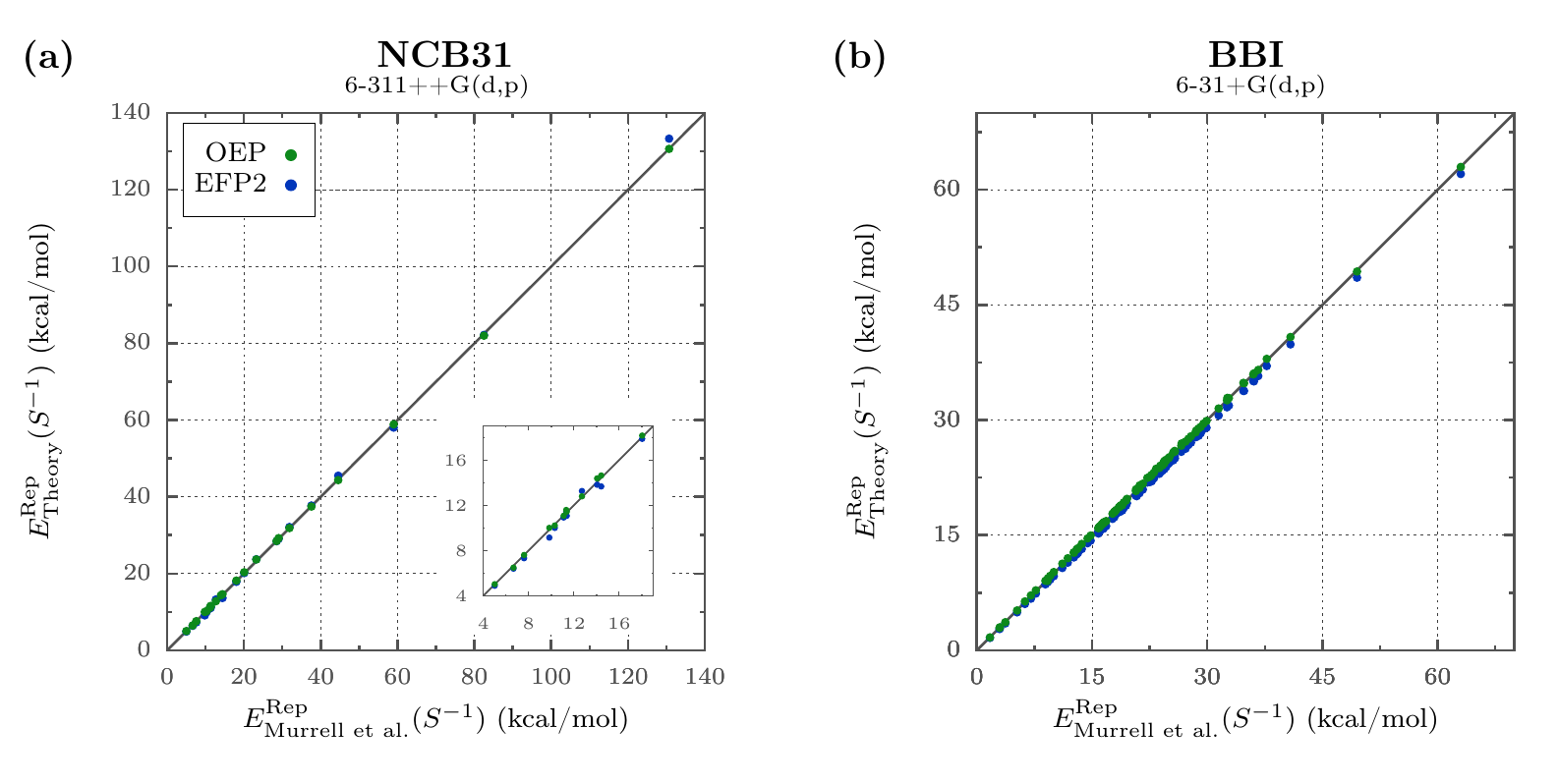}
\caption{\label{f:fig-2} {\bf Accuracy of the OEP and EFP2 models of first\hyp{}order repulsion energy
across various bi\hyp{}molecular systems.} 
(a) NCB31 
database\cite{Zhao.Schultz.Truhlar.JCTC.2006,
Zhao.Truhlar.JCTC.2005,Zhao.Schultz.Truhlar.JCTC.2006,Zhao.Schultz.Truhlar.JCP.2005} 
of non\hyp{}covalent interactions
and
(b) BBI subset\cite{Burns.Faver.Zheng.Marshall.Smith.Vanommeslaeghe.MacKerell.Merz.Sherrill.JCP.2017} 
of backbone\hyp{}backbone interactions in proteins from the BioFragment Database.
For the OEP calculations, the EDF-1 scheme with the aug-cc-pVDZ-jkfit auxiliary basis set
was used.
} 
\end{figure*}

\clearpage
\section{References}

\bibliography{references}